\documentclass[12pt,english,preprint]{emulateapj}
\usepackage[T1]{fontenc}
\usepackage[latin1]{inputenc}
\usepackage{amssymb}


\makeatother

\usepackage{babel}

\makeatother

\usepackage{babel}

\begin{document}

\title{Binary planetesimals and their role in planet formation}

\author{Hagai B. Perets}

\email{hperets@cfa.harvard.edu}

\affil{Harvard-Smithsonian Center for Astrophysics, 60 Garden St.; Cambridge,
MA, USA 02138}
\begin{abstract}
One of the main evolutionary stages of planet formation is the dynamical
evolution of planetesimal disks. These disks are thought to evolve
through gravitational encounters and physical collisions between single
planetesimals. In recent years, many \textit{binary} planetesimals
have been observed in the Solar system, indicating that the binarity
of planetesimals is high. However, current studies of planetesimal
disks formation and evolution do not account for the role of binaries.
Here we point out that gravitational encounters of binary planetesimals
can have an important role in the evolution of planetesimal disks.
Binary planetesimals catalyze close encounters between planetesimals,
and can strongly enhance their collision rate. Binaries may also serve
as additional heating source of the planetesimal disk, through the
exchange of the binaries gravitational potential energy into the kinetic
energy of planetesimals in the disk. 
\end{abstract}


\section{Introduction}

Current models for the evolution of planetesimal disks consider three
basic processes: viscous stirring, dynamical friction and coagulation
or disruption through collisions \citep[e.g. ][]{saf72,lis+93,gol+04}.
These processes account for the rate at which single planetesimals
encounter one another and collide, thereby changing the sizes (and
masses) of planetesimals and their velocity distribution. None of
these processes, however, takes into account the role of binary planetesimals.

An important question when discussing the role of binaries is whether
they exist at all in planetesimal disks during their evolution. Evidently,
in the Solar system many binary planetesimals (BPs) formed and survived
to this day \citep[see][for possible formation and evolutionary scenarios]{gol+02,wei+02,dur+04,fun+04,ric+06,lee+07,sch+08,per+09,nes+10}.
Large binary fractions ($>10$ \%) are currently observed both for
asteroids and trans-Neptunian objects (TNOs) in the Solar system \citep{nol+08b,ker+06,man+07,ric+06,wal+09}.
These are found to have separations of up to tens or hundreds times
the radii of their single planetesimal components (for asteroids and
TNOs, respectively; up to 0.1 the Hill radii for these objects). The
currently observed large fractions of BPs in the Solar-system therefore
serves as a basic motivation for our study. We assume that such BPs
population exists in a given planetesimal disk, and study its role
in the planetesimals evolution. We note that a variety of mechanisms
were suggested for the origin of BPs in the Solar system \citep{ric+06,nes+10},
but only some limited aspects of their evolution have been explored
(e.g. \citealp{per+09} and references therein). 

The role of binaries in gravitational systems have been studied in
depth in the context of stellar clusters \citep{hil75,spi+87,hut+92}.
In such systems stellar binaries are known to serve as important dynamical
heating source of the clusters (so called binary heating), which slows
the collapse of the cluster. Binary stars are also known to have a
major role in introducing stellar collisions and accelerate the growth
of massive stars; it is thought that most stellar collisions are the
result of encounters between binary and single (or other binary) stars.
Here we suggest that BPs can play a similar role in planetesimal disks.
The cross section for close encounters of binaries can be much larger
than that of single planetesimals, thereby increasing their collision
rates. The potential energy of close binaries could be exchanged into
kinetic energy of the planetesimals (and vice-versa), and therefore
serve as an additional heating (cooling) source for the planetesimal
disk. In the following we explore both the process of binary induced
collisions as well as binary heating.

\section{Binary encounters }

The gravitational interactions between binary and single planetesimals
can lead to various outcomes including binary disruptions, exchanges,
resonant encounters and physical collisions.

In binary-single encounters energy is exchanged between the internal
orbital energy of the binary, $E_{bin}$, and the kinetic energy of
the incoming perturber, $E_{k}$. These are given by $E_{bin}=Gm_{1}m_{2}/2a$,
where $m_{1},m_{2}$ are the binary-components masses and $a$ is
the binary mutual separation, and $E_{k}=0.5mv^{2}$, where $m$ is
the typical mass of the perturbing single planetesimals, and $v$
is the relative velocity between the planetesimals (typically of the
order of the velocity dispersion of the planetesimals). A binary is
termed soft if $|E_{bin}|/E_{k}<1$ and hard if $|E|/E_{k}>1$. On
average, hard binaries get harder following an encounter, and soft
binaries get softer (the so called Heggie's law; \citealp{heg75}).
Soft binaries rapidly evaporate through such encounters and have a
weak affect on the energy budget of the system. However, encounters
with hard binaries lead, on average, to the significant loss of orbital
energy from the binary, making the binary harder, following the scatter
of the perturbing object into higher velocity. Therefore such a process
leads to the increase of the planetesimals velocities and the dynamical
heating of the system. Alternatively, a physical collision may occur
between two or all of the planetesimals involved in the encounter.
In the following we discuss the role of both these possibilities.

\subsection{Binary induced collisions}

The collision rate of a planetesimals is given by \begin{equation}
\Gamma=n\sigma v,\label{eq:rate}\end{equation}
 where $n$ is the number density of planetesimals in the disk, $\sigma$
is the cross section for the collision and $v$ is the relative velocity
between the planetesimals. The cross section for a physical collision
depends on the relative velocity between particles. When the relative
velocity between planetesimals is much larger than the escape velocity
of the most massive planetesimal participating in the encounter, gravitational
focusing is negligible and the cross section is given by the projected
cross sectional of the planetesimal $\sigma\sim\pi r^{2}$ ($r$ is
the planetesimal radius). When the relative velocity is slower, gravitational
focusing becomes dominant and we obtain the following cross section
($\sigma_{1}$) for two single planetesimal collisions \begin{equation}
\sigma_{1}\thickapprox\pi r^{2}\left(1+\frac{v_{e}}{v}\right)^{2}\sim\pi r^{2}\left(\frac{v_{e}}{v}\right)^{2},\label{eq:cross section}\end{equation}
 where $v$ is the relative velocity and $v_{e}$ is the escape velocity
from the planetesimal surface (with the right most term obtained for
$v_{e}\gg v$) .

Binary-single encounters could involve complex trajectories, as the
encounter now involves three bodies and the planetesimal trajectories
can become chaotic (binary-binary encounters, not discussed here,
interact in even more complex way). In such encounters the probability
for a direct collision between any two (or even all) of the objects
involved is highly increased \citep{hil+80,fre+04}, and therefore
binaries serve as efficient catalysts for direct physical collisions.
The cross section for a physical collision during a single-binary
encounters, $\sigma_{2},$ is approximately given by \citeauthor{fre+04}
(2004; also Fregeau, private communication and \citep{sig+93,val+06}
and references therein) \begin{equation}
\sigma_{2}\approx\pi a^{2}\left(\frac{v_{c}}{v}\right)^{2}\left(\frac{r/10\, km}{a/2.14\times10^{3}km}\right)^{0.65},\label{eq:sigma-collision-in-binary}\end{equation}
 where $v_{c}$ is the critical velocity separating soft and hard
binaries and $a$ is the binary semi-major axis. Note that this equation
was derived for stellar encounters, however, the dynamics of gravitational
encounters is scale free, and can be scaled for the use of planetesimals
mass objects. Although physical collisions also depend on the density
of the objects, the typical average density of planetesimals ($0.5-3$
gr cm$^{-3}$) is comparable to that of stars, and therefore scaling
of mass and radii could be used. 

For simplicity we will consider equal mass planetesimals. For this
case $v_{c}$ is approximately the mutual orbital velocity of the
binary components (similar to the escape velocity from the binary
at its separation, $a$; i.e. $v_{c}\propto v_{e}(r/a)^{1/2}$) and
we therefore get \begin{equation}
\frac{\sigma_{2}}{\sigma_{1}}\approx33\left(\frac{a}{r}\right)^{0.35}.\label{eq:collision-sigma-ratio}\end{equation}

As can be seen from this ratio, collisions during binary-single encounters
could be tens up to hundreds of times more frequent than single-single
encounters for typical BPs currently observed in the solar system%
\footnote{Note that this ratio exceeds unity even as we approach to $r=a$;
in these cases the flyby of a perturber can easily perturb the binary
components into collision, even if the perturber never crosses between
the binary components. %
}. The ratio between collisions rates due to binary-single encounters
($\Gamma_{2}$), vs. single-single encounters ($\Gamma_{1}$) is therefore
given by

\begin{equation}
\frac{\Gamma_{2}}{\Gamma_{1}}\approx\frac{n_{2}\sigma_{2}v}{n_{1}\sigma_{1}v}\approx39\left(\frac{n_{2}/n_{1}}{0.3}\right)\left(\frac{a/r}{50}\right)^{0.35},\label{eq:collision-rate-ratio}\end{equation}
 where we took a binary fraction of $f_{bin}=n_{2}/n_{1}=0.3$ in
the normalization.

Consequently, the rate of physical collisions and the mass growth
of planetesimals (which is proportional to the collision rate, $\dot{M}\propto\Gamma$)
could be boosted and dominated by binary-single encounters even for
a low binary fraction. Moreover, binary-single encounters produce
a different (power law) dependence of the growth rate on the planetesimals
size. They also introduce dependencies on the \textit{binary} parameters,
namely the binaries separation and binary fraction (which could be
size dependent by themselves). The mass growth rate of planetesimals
could therefore change both quantitatively (faster) and qualitatively
due to existence of BPs. This could affect the size distribution of
planetesimals in protoplanetary disks (or debris disks) and its evolution.
Hence, the orbital properties of binaries \citep{nao+10}, the size
distribution of planetesimals, and their coupling could all be used,
in principle, to characterize and constrain the evolution of the Solar
system.

Taking a conservative binary fraction of 10 percents for binary asteroids
with typical separations of 10 times the planetesimal radius (e.g.
typical of main belt asteroids; \citealp{ric+06}), we get $\Gamma_{2}/\Gamma_{1}\sim7$.
The binary fraction of TNO binaries could be higher than 30 percents
with typical separations of $a\sim50\, r$ (e.g. typical for binary
TNOs; \citealp{ric+06}), leading to collision rate as high as $\sim39$
times higher than the expected single-single physical collision rate.
Given the likely higher binary fraction of planetesimals, both today
and in the past, collision rates catalyzed by BPs during planet formation
were likely to be even higher. Nevertheless, more detailed study of
the formation and destruction mechanisms of BPs is required to asses
this question quantitatively.

We note that in the case of single-single collisions, the planetesimals
are unbound prior to collision, whereas in the binary-single case
the colliding planetesimals could have been marginally bound (e.g.
during resonant encounters). The expected impact velocities during
binary-single encounters are therefore likely to be lower, on average,
than those expected in single-single collisions. Lower velocity collisions
are more likely to result in mass accretion rather than shattering
of planetesimals (the likely outcome of higher velocity impacts),
and are therefore more efficient for planetesimals growth. The accelerated
collision rate may also imply that more collisions occur at earlier
times, when the planetesimal disk is cooler, again leading to typically
lower velocity impacts.

\subsection{Binary heating}

In the following we focus on the more energetically important hard-binaries.
Encounters with hard binaries lead, on average, to the loss of orbital
energy from the binary and the scattering of the perturbing object
into higher velocity. Therefore such a process leads to the heating
of the system \citep{heg75,hil75,spi+87}. In fact, in stellar clusters,
this {}``binary heating'' process is considered as one of the main
processes governing the evolution of the clusters \citep{hut+92}.
We can now consider the effect of binary heating on the evolution
of a planetesimal disk. We first note, however, that even the initial
formation of binaries would heat the planetesimals disk (or the gas
if the binaries formed due to gas interactions), since the binaries
binding energy have had to be transferred to the disk planetesimals
upon their formation.

\subsubsection{Energy budget}

One can estimate the amount of potential energy reservoir available
for hard binaries. A binary can be hardened up to the point when it
becomes a contact binary. The potential energy of such binary is of
the order of \begin{equation}
E_{bin}\approx\frac{Gm^{2}}{2r}\label{eq:Ebin}\end{equation}
 where $r$ is the typical radius of a planetesimal of mass $m$.
The energy extracted from a binary, which was initially at some typical
separation $a\gg r$ is of the order of \begin{equation}
\Delta E_{bin}=E_{fin}-E_{init}=\frac{Gm^{2}}{2}\left(\frac{1}{r}-\frac{1}{a}\right)\approx\frac{Gm^{2}}{2r}=E_{bin}.\label{eq:Delta-Ebin}\end{equation}
 If all this energy were to be gained by the planetesimals in the
disk, they should have been excited to higher velocity dispersion,
with \begin{equation}
\Delta E_{kin}=N\frac{m\Delta v^{2}}{2}\approx N_{bin}\frac{Gm^{2}}{2r}=N_{bin}E_{bin},\label{eq:energy exchange}\end{equation}
 where $\Delta v$ is the change in the velocity dispersion of the
planetesimals, and $N$ and $N_{bin}$ are the number of single and
binary planetesimals, respectively. We therefore find that the change
in velocity is \begin{equation}
\Delta v\approx\sqrt{G\frac{N_{bin}m}{Nr}}=\sqrt{f_{bin}\frac{Gm}{r}},\label{eq:scatter v}\end{equation}
 where $f_{bin}$ is the binary fraction. Comparing this to the Hill
velocity, $v_{H}=(Gm/R_{H})^{1/2}$ (where $R_{H}\simeq A(3m/M_{\odot})^{1/3}$
is the Hill radius at distance $A$ from the Sun), we find that \begin{equation}
\Delta v\sim\sqrt{f_{bin}\frac{R_{H}}{r}}v_{H}.\label{eq:scatter-v vs. vH}\end{equation}

Typically, the radii of planetesimals are much smaller than the Hill
radius by a few orders of magnitude. Therefore, even a small binary
fraction potentially holds enough energy to heat up a disk to Hill
and even super-Hill velocities. This could have important implications
on the binary formation mechanism involved. For example, binary formation
through dynamical friction discussed by \citet{gol+02}, is efficient
only for relatively low velocity dispersions of the small planetesimals.
This may imply that the high binary frequency observed in the Solar
system is limited to large planetesimals; e.g. the energy extracted
from the binaries formation is distributed over a much larger mass
of low mass planetesimals, which never formed binaries by themselves
as these were all too soft to survive and contribute to the binary
heating. Alternatively, binary formation could have been primordial,
when they were embedded in gas, or during the collapse of planetesimals
swarms forming single planetesimals \citep{nes+10}. In these latter
cases the extra energy from the binaries formation would be dissipated
in the gas or radiated as heat during the gravitational collapse.
Better understanding of binaries formation is required for resolving
this issue.

\subsubsection{Heating rate}

We now turn to the rate at which binaries heat the system. For simplicity
we will consider a planetesimal disk composed of two types of objects:
single planetesimals of mass $m$, number density $n$, and typical
velocity dispersion $v$ and hard binaries of mass $m_{bin}$ (where
we assume the mass ratio, $q$, is typically high), number density
$n_{bin}$. The rate of energy gain by the system from a single hard
binary is then given by \citep{hil+80,spi+87,hil92}

\begin{equation}
\frac{dE_{bin}}{dt}=D_{2}G^{2}\frac{nmm_{bin}^{2}}{v{}_{enc}},\label{eq:dEbin}\end{equation}
 where $n$ is the number density of planetesimals, $v_{enc}$ is
the typical encounter velocity ($\sim$ the velocity dispersion of
the planetesimals), and $D_{2}$ is some constant pre-factor, which
will be discussed later on. We now use the relation $dE/dt=3mvdv/dt$
(the 3 factor comes from considering the velocity dispersion in a
3D system) to find expressions for the binary heating of the single
planetesimals. Taking \begin{equation}
n\frac{dE}{dt}=-n_{bin}\frac{dE_{bin}}{dt}\label{eq:Eeq}\end{equation}
 and translating this into evolution of the velocity dispersion (i.e.
the energy extracted from the binaries is the kinetic energy gained
by the planetesimals), we find

\begin{equation}
(\frac{dv}{dt})_{bin}=\frac{n_{bin}}{n}\frac{1}{3mv}\frac{dE_{bin}}{dt}=D_{2}G^{2}\frac{n_{bin}m_{bin}^{2}}{3vv_{enc}}\approxeq D_{2}G^{2}\frac{n_{bin}m_{bin}^{2}}{3v^{2}},\label{eq:dv}\end{equation}
 where numerical simulations \citep{hil92} show that the constant
pre-factor, $D_{2}$, is of the order $6.7$ for encounters of single
mass objects (i.e. $q\sim1$ and $m_{bin}=2m$) and about twice as
large for the case of more massive binaries ($m\ll m_{bin}$). In
the last term we assume the encounter velocity of of the same order
of the velocity dispersion of the planetesimals. 

We can compare the importance of binary heating with that of viscous
heating by single planetesimals. Let us use a simple estimate for
the viscous heating, following a similar approach by \citet{ale+07},
taking a simplifying assumption that the velocity dispersion is isotropic.
This is not strictly valid, but it has been shown that the ratio of
the radial and vertical velocity dispersions cannot become larger
than 3 without the system becoming unstable \citep{kul+71,pol+77}.
The relaxation time for such a system is given by \begin{equation}
t_{\mathrm{relax}}=\frac{v^{3}}{CG^{2}nm^{2}\ln\Lambda}\,,\label{eq:t_rel}\end{equation}
 (e.g. \citealp{bin+87,pap+01} where $n$ is the planetesimals density,
$v$ is the typical velocity dispersion of the planetesimals and $\ln\Lambda$
is the Coulomb logarithm (where $\Lambda\sim H/r$; H is the disk
scale height) and $C$ is an order-of-unity constant that depends
on the geometry of the system. (for a spherical system $C\simeq2.94$,
\citealp{bin+87}). Consequently, the relaxation of the system is
governed by\begin{equation}
\left(\frac{dv}{dt}\right)_{1}=D_{1}G^{2}\frac{nm^{2}\ln\Lambda}{3v^{2}},\label{eq:viscous-heating}\end{equation}
with $D_{1}=2C.$ Comparing with the binary heating term in Eq. \ref{eq:dv}
we find both heating mechanism have similar dependencies on the disk
and planetesimals properties. Taking the same velocity dispersion
for both binaries and single planetesimals, and considering the same
mass for all planetesimals ($m_{bin}=2m$), the two heating terms
differ only by some constant pre-factor, and the relative binary fraction,
i.e.\begin{equation}
\left(\frac{dv}{dt}\right)_{bin}/\left(\frac{dv}{dt}\right)_{1}=\frac{4D_{2}}{D_{1}}\frac{n_{bin}}{n}=\frac{4D_{2}}{D_{1}}f_{bin}.\label{eq:bin0-single-ratio}\end{equation}
Taken at face value $4D_{2}/D_{1}\sim4.5$, and therefore the binary
heating contribution could be comparable to that of viscous heating
and dynamical friction, for realistic binary fractions of $0.1-0.2$.
Note that in the case of dynamical friction effect on low mass planetesimals,
we should take the binary mass to be larger than the mass of the heated
planetesimals and $D_{2}$ would be twice as large \citep{hil92}.
Given these uncertainties and the simplifications used in the above
derivations, one should be cautious in using them at face value. These
findings do, however, suggest that binary heating is likely not negligible,
but is also not likely to be the single most dominant mechanism for
heating of the planetesimals disk. 

Binary heating would continue as long as hard binaries exist in the
disk. However, as with the other forms of planetsimal disk heating,
heating the disk puffs it up, resulting in a larger volume and hence
lower number density of planetesimals and lower encounter rates. 

In addition, as the velocity dispersion is increased, more binaries
become soft, and could be disrupted. We note that all of the the suggested
binary origin scenarios become less efficient at higher velocity dispersions,
as the number density of planetesimals decrease and gravitational
focusing becomes less effective. Currently, most of the observed BPs
are soft in terms of encounters with similar size/mass planetesimals,
and none of the binary planetesimals formation scenarios is currently
effective in producing new binaries (\citealp{ric+06}; beside the
formation of low mass binaries through radiative spin up in the inner
regions of the Solar system). Therefore binary formation and major
collisional evolution had to proceed at early times, and the observed
binary fraction today is representative only of the survived binaries,
where as the earlier binary fraction was likely to be higher.

\section{Caveats}

\textbf{Stellar vs. planetesimal encounters: }The main caveat in our
discussion of the role of BPs is the use of the stellar encounters
approximation for estimating planetesimals encounters. Such approximations
neglects the effect of the Sun on the encounters. 

The important regime for binary-single encounters is when the impact
parameter of the encounter is smaller than the binary separation (and
the Hill radius of the binary). Such encounters therefore occur mainly
in the regime dominated by the mass of the planetesimals rather than
the Sun. However, during resonant encounters, planetesimals can be
scattered to distances larger than the Hill radius, and still come
back to re-encounter the binary in the absence of the Sun potential,
at which point the tidal forces induced by the Sun could perturb them.
Most resonant orbits, however, are likely to be at smaller separations
(see e.g. \citealp{hil83}, for a related, although different problem);
in fact the probability of being ejected to some large distance $r\gg a$
is comparable to the probability of the binary becoming unbound altogether
(see \citealp{val+06}, chapter 8.3). Such perturbation and possible
quenching of the interaction due to the Sun would therefore mostly
affect wide binaries, with separations close to their Hill radius.
Note, however, that when taking into account the gravitational pull
of the Sun, the interaction even between single planetesimals could
become resonant, basically leading to the temporary capture of the
planetesimals and a long lived interaction \citep{ast+05,raf+10}.
The additional interaction with the Sun can therefore increase the
rate of resonant encounters in which the the BPs interact chaotically
and therefore have a higher probability for physical collisions. Few
body simulation of such binary-single planetesimals, which are beyond
the scope of this letter could give a more quantitative picture of
these processes. 

\textbf{Collisional cooling: }In the discussion of binary heating
we neglected the effect of physical collisions. The amount of energy
damping through collisions is not very well understood for planetesimals
and depend on the (unknown) coefficient of restitution of the planetesimals.
The high collision rate during encounters could therefore serve an
important role in the heating/cooling rate by binaries. We note that
the importance of collisions is poorly known even in the much better
studied literature of binaries in stellar clusters, where it was suggested
to play an important role mostly for the hardest (tidally captured)
binaries \citep{mcm86} making binary heating much less efficient.

\section{Discussion and summary}

In this \emph{letter} we pointed out the importance of BPs and their
role in the evolution of protoplanetary and debris disks. They could
affect the evolution of such disks both through efficiently catalyzing
physical collisions between planetesimals, and by serving as an additional
planetesimal heating/cooling mechanism.

Binary encounters provide a novel and efficient mechanism for planetesimals
growth, suggesting a different dependence between the growth rate
of planetesimals and their physical size. They also introduce dependencies
on the binary properties, such as the binary fraction and semi-major
axis (and their size dependent distribution). The evolution of the
planetesimals size distribution in this case could therefore be qualitatively
different than that envisioned in studies taking into account only
single planetesimals \citep[e.g.][and other related papers and references therein]{doh69,dav+97,ken+04}{]}.

The relative importance of binaries depends on the binary fraction
of planetesimals. Current observations of the Solar system show large
BPs (asteroids; TNOs) populations, even amongst the largest planetesimals/embryos,
e.g. Pluto-Charon and the Earth-moon systems. BPs therefore exist
even at late stages in the planetesimal disk, even up to the the scale
of planetary embryos. BPs may therefore accelerate the mass growth
of planetesimals both at early stages of evolution (when they were
suggested to form; \citealp{gol+02}), and possibly even up to the
stages relevant for the formation of the gas planets cores.

We note that approaches used for the study of binaries in stellar
clusters, could similarly be used to confront the new challenges raised
by accounting for the role BPs, but we caution, and raise some caveats
for their direct use in this context. 

Current studies of planet formation do not take into account BPs.
Such additional component could be difficult to account for in these
studies (especially in simulations where following binary orbits is
computationally expensive). Nevertheless, our findings suggest that
not only that binaries are not negligible, but they may have an important
role in the evolution of planetesimal disks, and therefore should
not be ignored. We conclude that the inclusion of binaries in future
studies could have important implications for our understating of
the evolution of planetesimal disks and planet formation.

\acknowledgements{The author thanks John Fregeau, Scott Kenyon, Ruth Murray-Clay, Noam Soker and the anonymous referee for helpful discussions/comments. The
author is a CfA, BIKURA(FIRST), Fulbright and Rothschild fellow.}

\bibliographystyle{apj}

\end{document}